\RequirePackage{fix-cm}
\documentclass[smallextended]{svjour3}       
\smartqed  
\usepackage{graphicx}
\usepackage{amsbsy}
\usepackage{amssymb}
\usepackage{amsmath}
\usepackage{cite}
\usepackage{xcolor}
\usepackage{caption}
\usepackage[labelformat=simple]{subcaption}

\newcommand*{\B}[1]{\ifmmode\pmb{#1}\else\textbf{#1}\fi}
\newcommand*{\et}{\textit{et al}}
\newcommand{\red}[1]{\textcolor{black}{#1}}
\newcommand{\blue}[1]{\textcolor{black}{#1}}
\begin{document}

\title{Dynamics of fine particles due to quantized vortices on the surface of superfluid $^4$He
}


\author{Sosuke Inui        \and
  Makoto Tsubota      \and
   Peter Moroshkin    \and
   Paul Leiderer        \and
   Kimitoshi Kono.
}

\authorrunning{Short form of author list} 

\institute{S. Inui \at
              Osaka City University, Osaka, Japan \\
              Tel.: +81-66-605-2530\\
              \email{sinui@sci.osaka-cu.ac.jp}           
           \and
           M. Tsubota \at
           Osaka City University, Osaka, Japan\\
           \and
           P. Moroshkin\at
           Okinawa Institute of Science and Technology, Okinawa, Japan\\
           \and
           P. Leiderer\at
           University of Konstanz, Konstanz, Germany\\
	 	  \and	           
           K. Kono\at
           National Chiao Tung University, Taiwan\\
}

\date{Received: date / Accepted: date}

\maketitle

\begin{abstract}
We have conducted calculations of the coupled dynamics of \red{quantized} vortices and fine \blue{metallic} particles trapped at a free surface of superfluid $^4$He.  The computational result so far indicates that a particle-vortex complex may produce \red{quasi-periodic} motions  \red{along the surface} and that the motions can be enhanced if the metallic particles are heated and induce local radial flows.  \red{Our results qualitatively reproduce recent experimental observations of trapped particle motion.} 

\end{abstract}

\section{Introduction}
\label{intro}
A new dynamical aspect of superfluid $^4$He that involves both bulk quantized vortices and the \red{free} surface of the system may have been revealed by the experiment utilizing a number of electrically charged fine metallic particles trapped just below the surface of superfluid $^4$He by Moroshkin \et.  They produce the particles by a laser ablation of a metallic target  \red{submerged} in superfluid $^4$He \cite{Moroshkin_2010,Moroshkin_2016}, and trap them just below the \red{free} surface \red{by using a combination of surface tension and the applied static electric field.}  The particles are illuminated with \red{the expanded beam of} \blue{a cw} laser, and their motions are tracked.  Most of \red{the particles} rest at their equilibrium positions; however, some \red{of them} are observed to exhibit anomalous quasi-periodic motions that are largely classified into two types; (1) circular motions with specific frequencies \red{and radii}, and (2) quasi-linear oscillations with sharp turning points. \cite{Moroshkin_LT}  

The \red{injection} of fine particles  into superfluid $^4$He is a method that has been appreciated for \red{decades} for various purposes and has achieved a big success in various cases\cite{Bewley_2006,Sergeev_2009,Guo_2014}, \red{but first observations of particles at a free surface have been reported only recently \cite{Moroshkin_2017}.}  We discuss that the motion of type (1) can be caused by a relatively simple \blue{vortex}-particle configuration\blue{:} a vertical vortex filament whose edges are terminated at the bottom of the vessel and at the particle just below the surface of $^4$He.  In \red{a} number of experiments with $^4$He$^+$ ions  (or ``snow balls") and negative ions (or ``electron bubbles") at the surface of superfluid $^4$He ever conducted\cite{Vinen_1995,Tabbert_1997}, the particles of type (1) and (2) were not observed either, yet these experimental set-ups and the one used by Moroshkin \et. have similarities.   However, there is a decisive difference.  That is that the illuminated metallic particles may act as local ``heaters'' in the experiment by Moroshkin \et., and it \red{can be} estimated that a particle can \red{be substantially heated by the intense laser beam and} create a local radial counter flow \red{with a velocity at the particle surface of order of 30 cm/s.}
 
\section{Method}
\label{sec:CM}
The anomalous motions of type (1) and (2) are presumably due to the interactions between quantized vortices and particles.  Thus, the coupled dynamics of quantized vortices and particles has to be considered.  That calculation is carried out based on the vortex filament model \cite{Schwarz_1988, Adachi_2010,Schwarz_1985}.  In this model, vortices are treated as \blue{very thin filaments}, and they are discretized into finite amount of segments. \blue{Each segment is represented by a point} $\B{s}(\xi)$, where $\xi$ is the arc length parameterization along the vortices.  The distance between the \red{neighbouring} vortex points determines the vortex resolution $\Delta \xi$.  In order to take account for the particle-vortex interaction, we introduce the equation of motion for a trapped particle as described in Ref. \cite{Mineda_2013}.  By making three assumptions, we are able to let a vortex point represent the trapped particle.  The assumptions are: (i) the particle radius is much smaller than the vortex resolution, (ii) a vortex reconnects with the center of the particle, and (iii) the particle and the vortex segment trapping it, in average, travel with the same velocity.
\begin{figure}
\centering
 \begin{subfigure}[b]{0.3\textwidth}
            \centering
            \includegraphics[width=\textwidth]{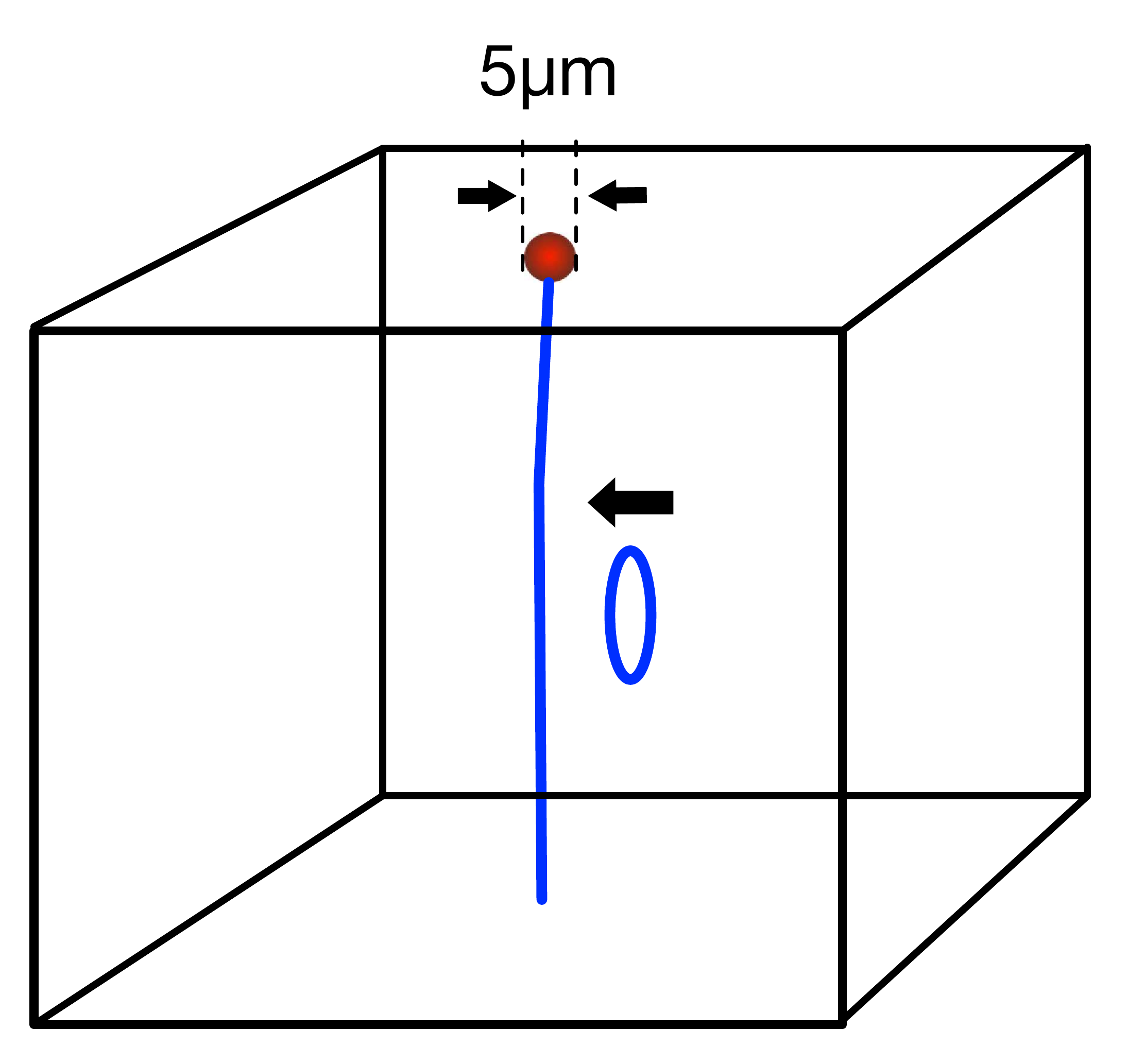}
    			\caption{Initial configuration\label{fig:Initial_Config}}
    \end{subfigure}
    \begin{subfigure}[b]{0.35\textwidth}
            \centering
            \includegraphics[width=\textwidth]{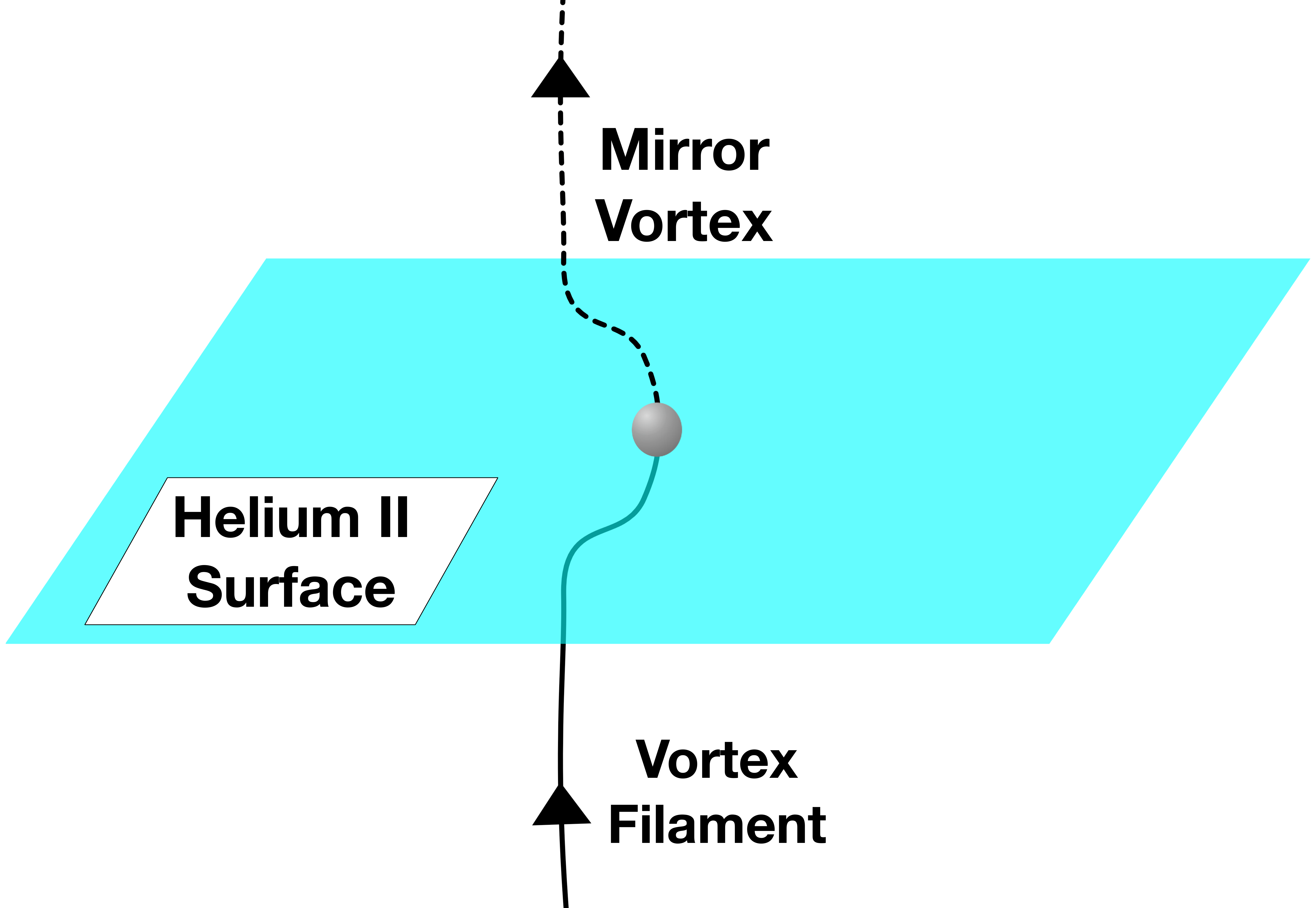}
    		   \caption{Zoomed in on the particle\label{fig:Zoom_in}}
    \end{subfigure}
    \caption {\blue{(a)} Illustration of the initial configuration of the system.  The top of the box corresponds to the $^4$He surface, and a particle of radius $5$ $\mu$m (a red ball) is trapped at the top of a straight vertical vortex filament (a blue line).  We excite the vertical filament by letting a small vortex ring (a blue ring) collide with it.  \blue{(b)} The \blue{trapped} particle \blue{at the surface} after the excitation. Only the tip of the vortex bends and starts to oscillate.}
\label{fig:Vor_Config}       
\end{figure}

The initial vortex-particle configuration considered in this set of calculations is shown in Fig. \ref{fig:Initial_Config}.  A straight vertical vortex filament trapping a particle on its top end sits still at the center of the system, and we excite the filament to see their time evolutions \red{by letting a small vortex ring collide with it}.  Here, the particle motion is confined to the two-dimensional $^4$He surface.  In order to reproduce the phenomena found in the experiment, we give suitable values to the parameters;  the system is  a cubic box of length $2$ mm, and a particle of radius a few $\mu$m is trapped at the top of the box filled with $^4$He.  At the top and the bottom surfaces we assume that the superflow is subject to the solid boundary condition.  The vortex resolution $\Delta \xi$ is set to be $10$ $\mu$m, which is larger than the particle radius, and the temperature is set to be $1.9$ K.
\begin{figure}
\centering
 \begin{subfigure}[b]{0.35\textwidth}
            \centering
            \includegraphics[width=\textwidth]{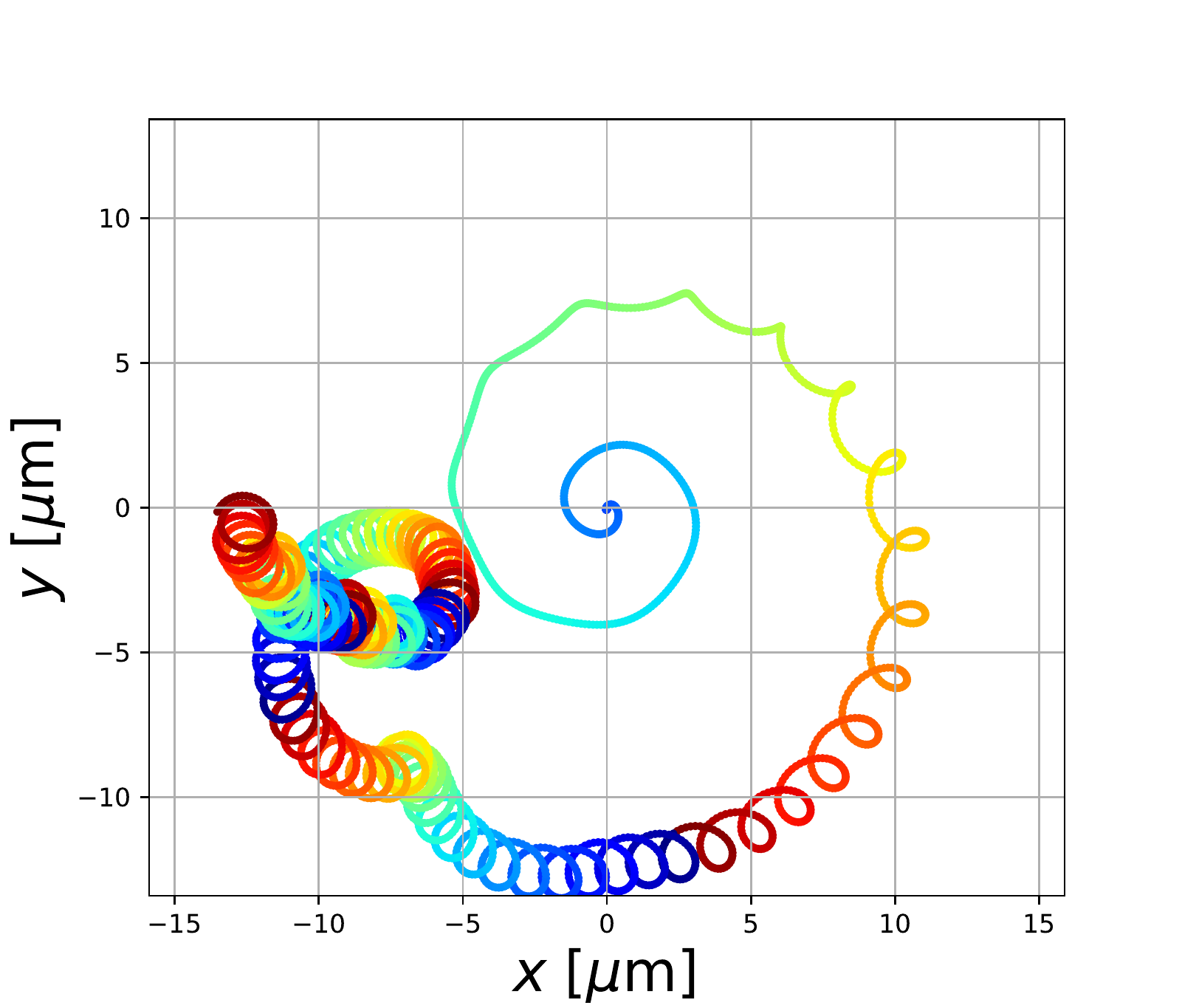}
    			\caption{Top view}
    \end{subfigure}
    \begin{subfigure}[b]{0.4\textwidth}
            \centering
            \includegraphics[width=\textwidth]{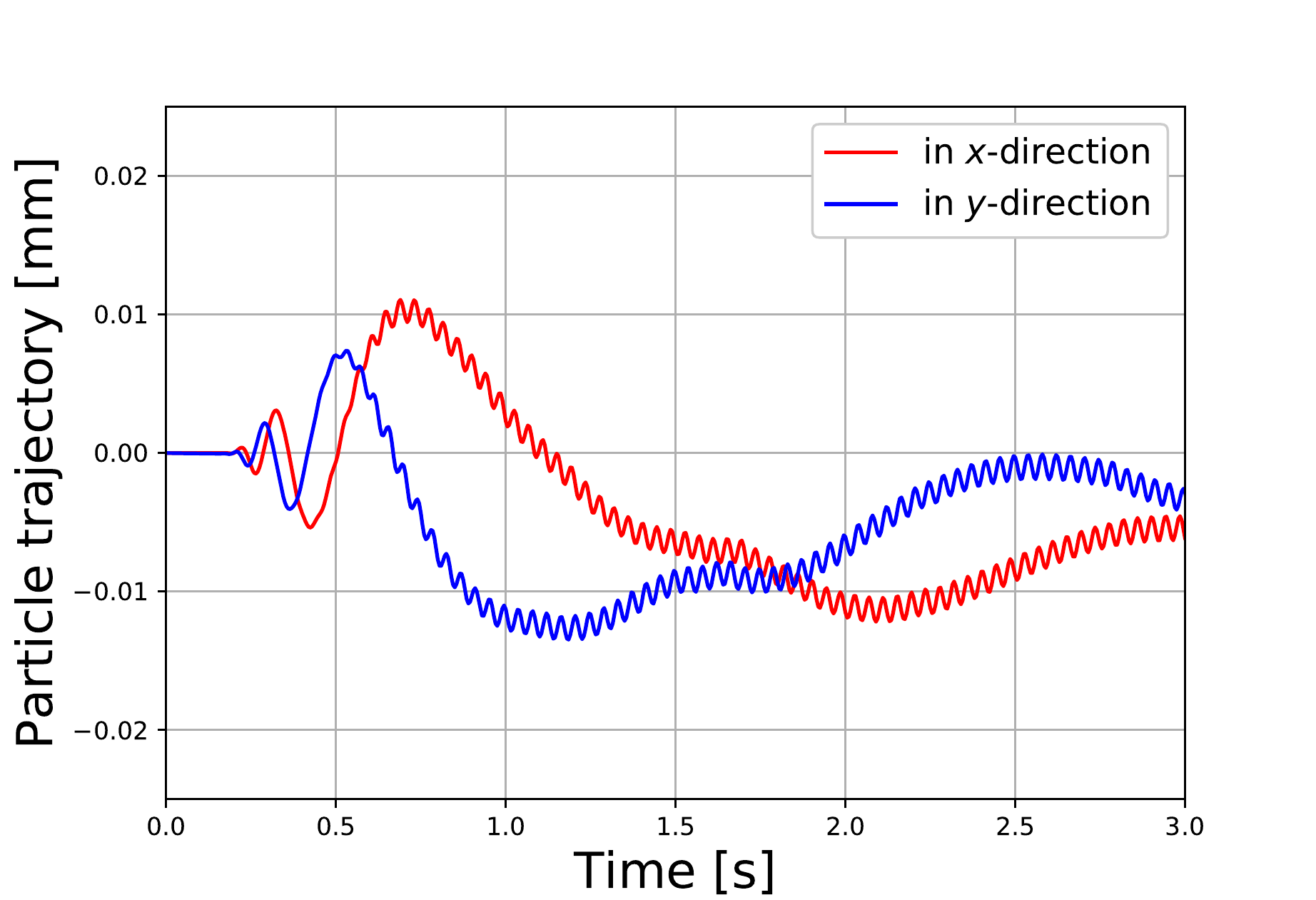}
    		   \caption{In $x$, $y$ -directions}
    \end{subfigure}
    \caption{A typical particle trajectory on the surface.   At the beginning it shows a large scale motion, but it decays quickly as a small scale circular motion of period \blue{$T_{traj}$} and radius \blue{$R_{traj}$} appears at around $0.6$ \blue{s} for this plot. }
\label{fig:ITypical_behaviour}       
\end{figure}

\section{Results and Discussion}
\label{sec:RD}
A typical motion of a particle trapped by a vertical vortex is shown in Fig. \ref{fig:ITypical_behaviour}. The particle is initially at rest at the origin which is an equilibrium position.  Once it is excited, the particle is kicked out of the origin and \red{starts to} spiral out.  However, the spiral motion decays as a small circular motion appears.  While the detail of spiral motion depends on the excitation, the small circular motion seems to be intrinsic for the particle-vortex complex.   We have examined the small circular motion by changing (i) the density and (ii) the volume of the particle.

\begin{figure}
\centering
 \begin{subfigure}[b]{0.4\textwidth}
            \centering
            \includegraphics[width=\textwidth]{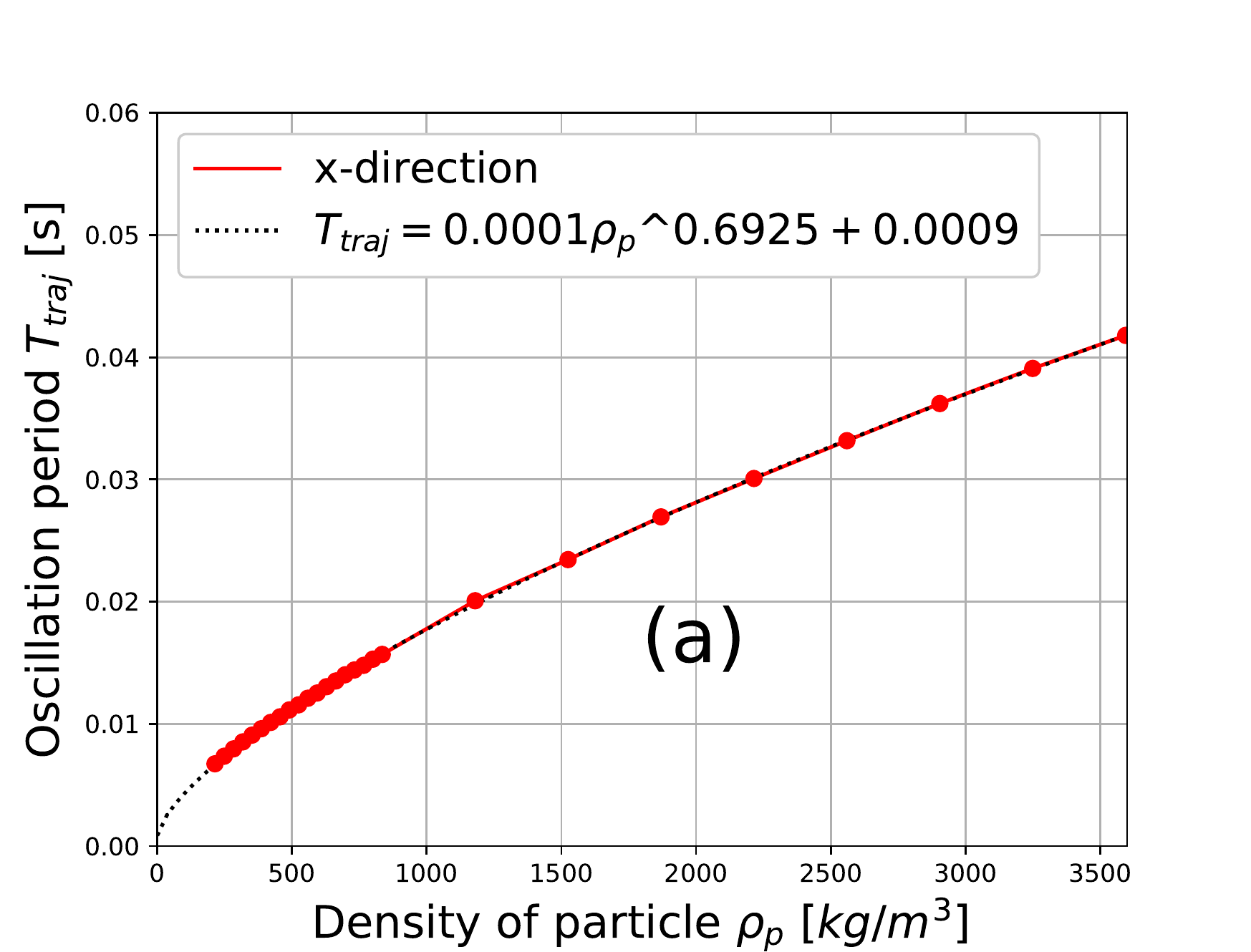}
    \phantomcaption
    \label{fig:a-dependence_a}
    \end{subfigure}
    \begin{subfigure}[b]{0.4\textwidth}
            \centering
            \includegraphics[width=\textwidth]{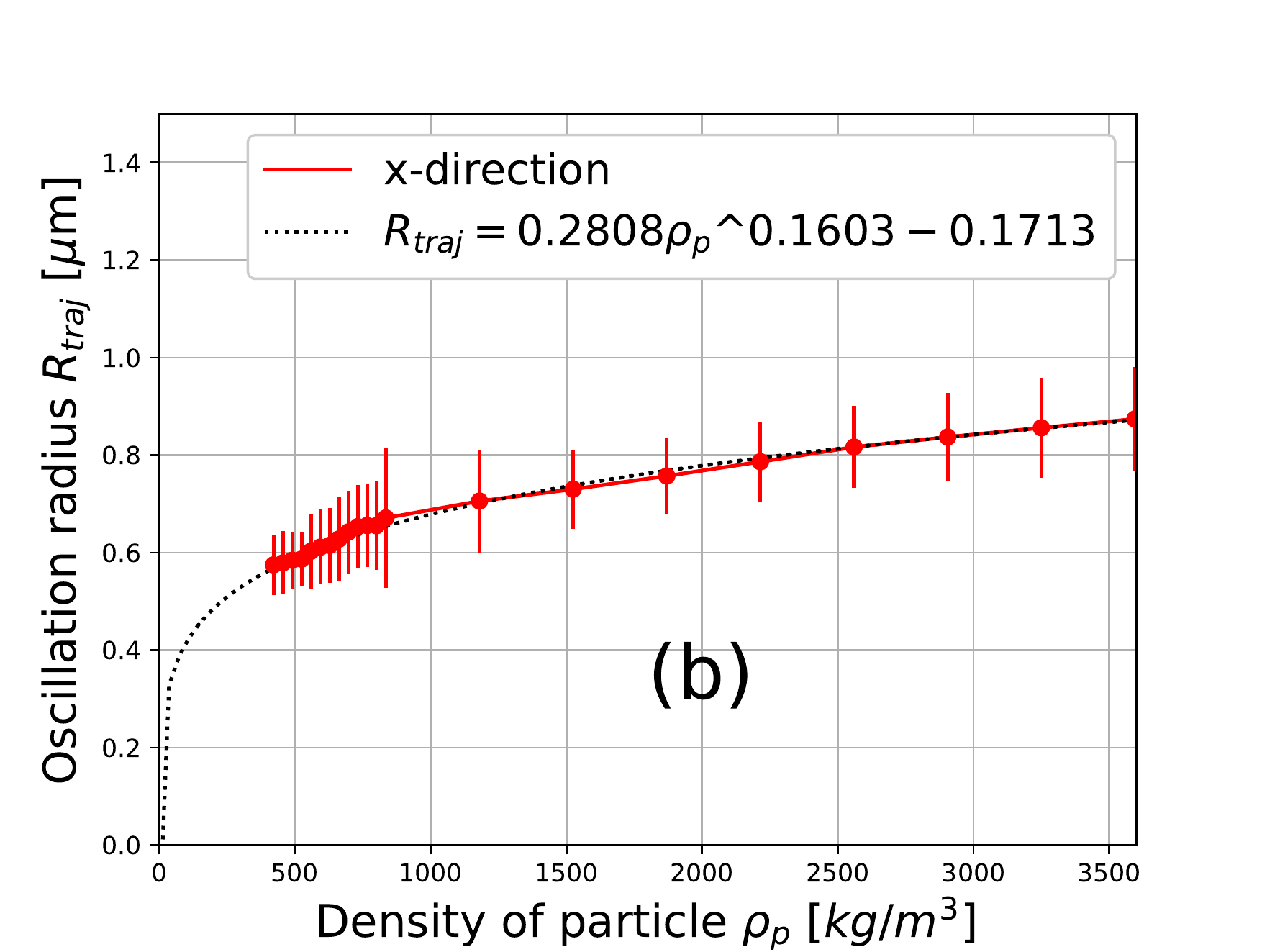}
    \phantomcaption
    \label{fig:a-dependence_b}
    \end{subfigure}
    \caption{(a) Period and (b) radius of the circular motion as a function of the particle density $\rho_p$.  Both \red{$T_{traj}$} and \red{$R_{traj}$} values are obtained by averaging over some time after the spiral motion decays sufficiently\red{, and error bars are obtained by taking the standard deviations}.  Since there are \blue{no} noticeable differences in the time-averaged motion in $x$ and $y$ -directions, each plot is represented by the motion in $x$-direction. Here, the particle \blue{radius $a_p$} is fixed to be $5$ $\mu$m. Both plots are fitted with a function of form $y=\alpha x^\beta + \gamma$ \blue{as shown in the legends} and represented by the black dashed curves.  $\rho_p = 3594.7$ kg/m$^3$ corresponds to the density of Ba at room temperature.}
\label{fig:a-dependence}       
\end{figure}
\begin{figure}
\centering
 \begin{subfigure}[b]{0.4\textwidth}
            \centering
            \includegraphics[width=\textwidth]{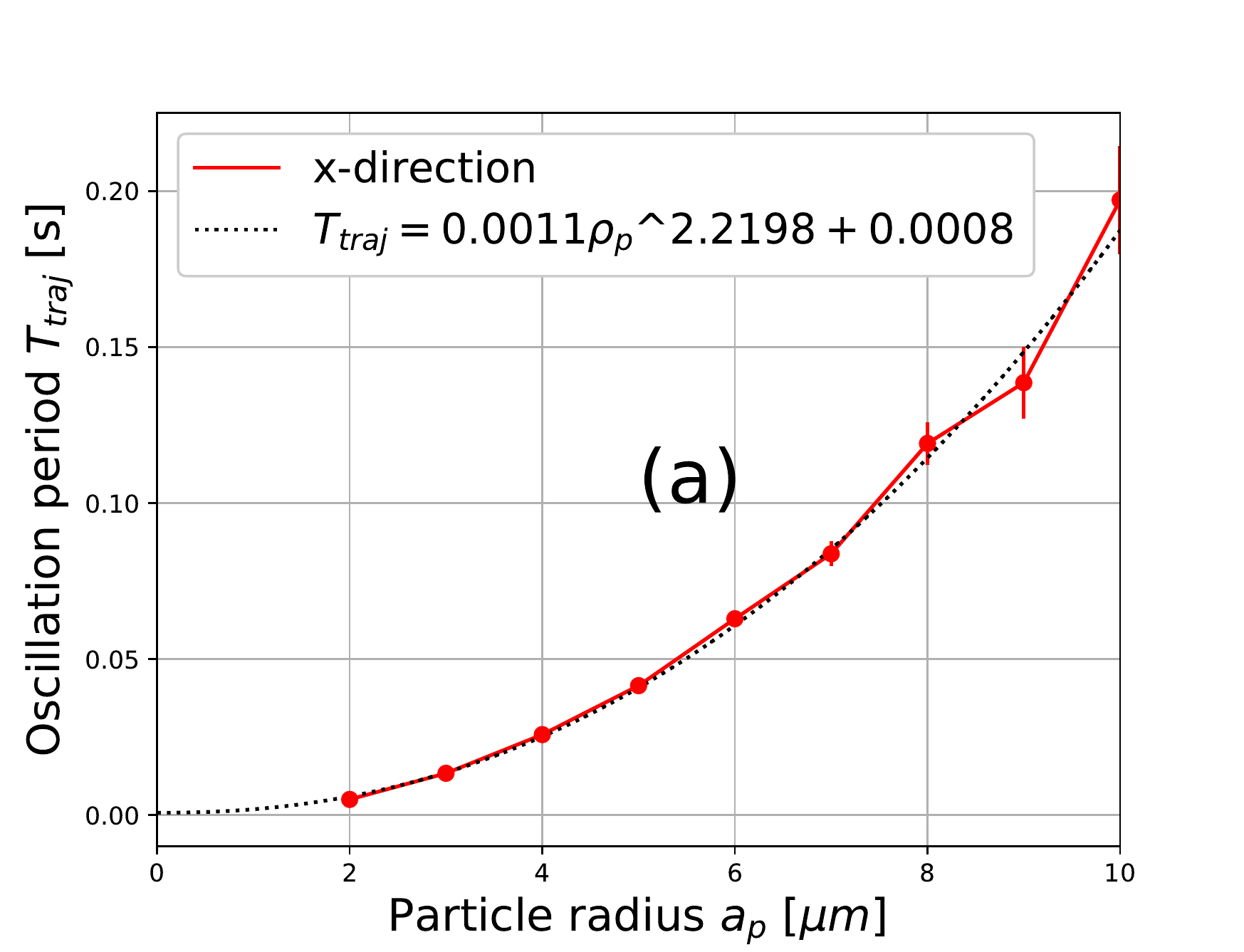}
    			\phantomcaption\label{fig:vol-dependence_a}
    \end{subfigure}
    \begin{subfigure}[b]{0.4\textwidth}
            \centering
            \includegraphics[width=\textwidth]{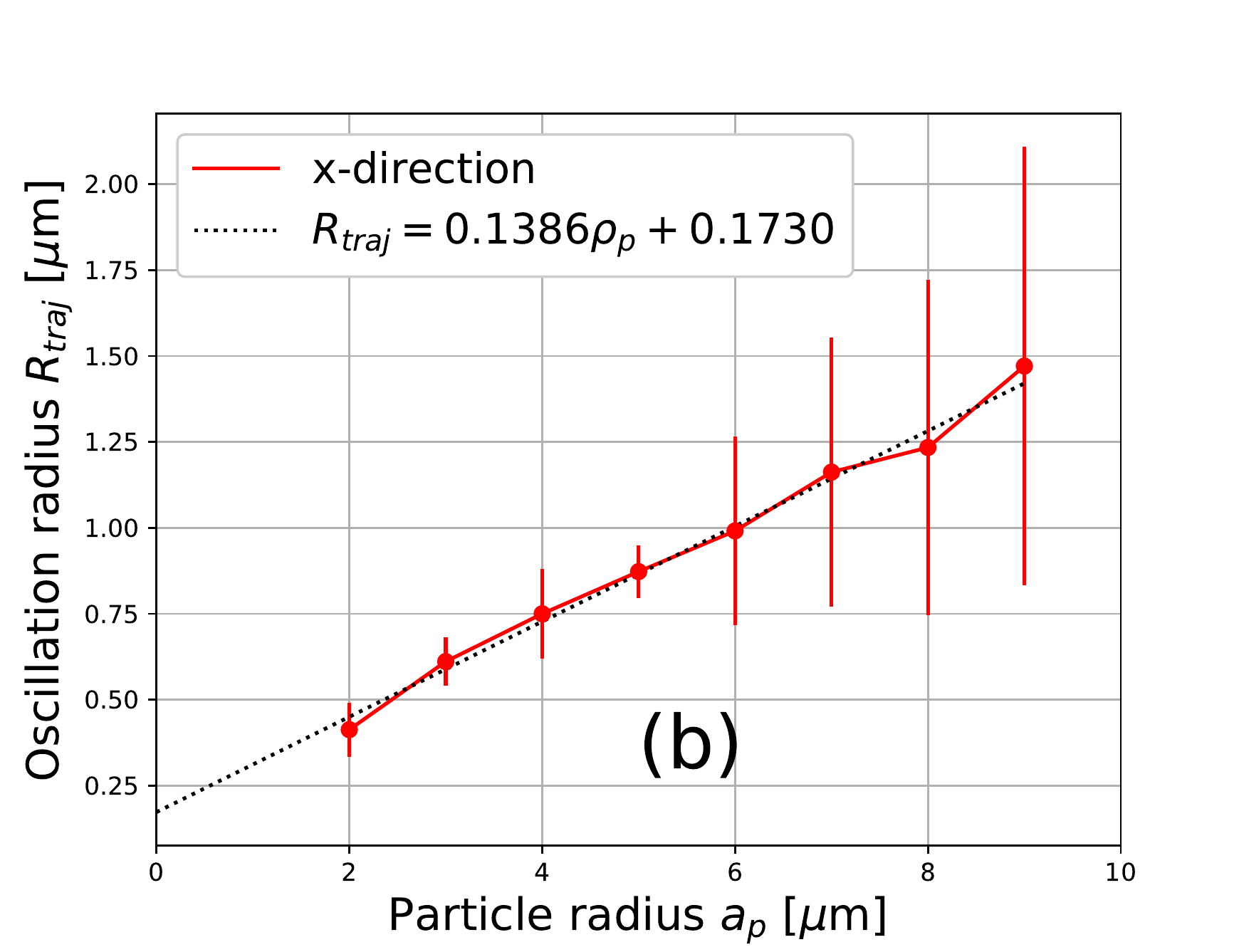}
    		   \phantomcaption\label{fig:vol-dependence_b}
    \end{subfigure}
    \caption{(a) Period \red{$T_{traj}$} and (b) radius \red{$R_{traj}$} of the circular motion as a function of the particle radius $a_p$.  The \red{$T_{traj}$} and \red{$R_{traj}$} values are obtained by time-averaging in the same way as in Fig.\ref{fig:a-dependence}, and the plots are represented by the motions in $x$-direction. The particle density is set to be that of \blue{barium}.  The plots are fitted with \blue{the functions shown in the legends}. }
\label{fig:vol-dependence}       
\end{figure}

It turns out that the period \red{$T_{traj}$} and trajectory radius \red{$R_{traj}$} of the intrinsic circular motion strongly depend on its density $\rho_p$ as shown in Fig. \ref{fig:a-dependence}.  Although the accurate asymptotic behaviours as $\rho_p$ approaches zero are not determined, we need to note that the fitted plot of Fig. \ref {fig:a-dependence_b} starts to drop quickly when $\rho_p \lesssim 200$ kg/m$^3$.  This indicates that it is difficult to detect the type (1) circular motions in the experiments with the \blue{light} particles such as solid hydrogen($\sim86$ kg/m$^3$) .  Not only the density but the volume of the particle also determines the motion.  The plots of the period and the radius of the circular motion as functions of the particle radius $a_p$ are shown in Fig. \ref{fig:vol-dependence}. 

\red{Although the Maryland group succeeded in visualizing a vortex filament by letting solid hydrogen particles be trapped onto it in series \cite{Bewley_2006}, meaning they have directly observed the particles trapped onto the vortices, the type (1) motions were not reported.  It seems that that is largely because solid hydrogen was used for the tracer particles.}
 In the experiment by Moroshkin \et., on the other hand, heavier metals such as Ba ($3594.7$ kg/m$^3$ at room temperature) are used, which \blue{makes it possible to observe} the new types of motions.  

Intuitively, the origin of the type (1) motion with this configuration can be qualitatively understood as \red{follows}: \red{We consider a straight vertical vortex filament trapping a particle at the surface. The vortex creates a concentric velocity field $\B{v}_{s,BS}$ that drops inversely proportional to the distance from the center.  We assume that the small displacements of the particle and the tip of the vortex trapping the particle \red{(see Fig.  \ref{fig:Zoom_in})} do not perturb the velocity profile significantly.  When the particle is out of the origin, it shows the circular motion about a new equilibrium location which is determined by balancing the two forces acting on the vortex segment; the Magnus force $\B{F}_M$ and the tension $\B{F}_T$. The Magnus force is given by }
\begin{equation}
 \B{F}_M \propto \B{s}^{\prime}\times \left[    \dot{\B{s}}-\B{v}_{s,BS}   \right],
\end{equation}
\red{where $\B{s}$ is the filament, and the prime and the dot symbols represent the derivatives with respect to arc length and time, respectively.  This force acts outward and is dominant when $\B{v}_{s,BS}$ is large.  On the other hand, the tension, given by}  
\begin{equation}
\B{F}_t \propto \B{s}^{\prime\prime},
\end{equation}
\red{only depends on the local curvature ($1/| \B{s}^{\prime\prime}|$ ) of the filament.  As the particle gets farther the local curvature of the vortex (and its mirror vortex) becomes smaller, and the tension force that pulls the particle backward becomes dominant.  Although we can qualitatively explain the origin of the type (1) motion, we need to note that the simulated radius of the trajectory is smaller by a factor of  more than $10$ than that found in the experiment, while the periods are in the same order.  This discrepancy may be explained by an effective radius mentioned in Sec.\ref{sec:EoHP}.}


\section{Effects of a heated particle}
\label{sec:EoHP}
\begin{figure}
\centering
 \begin{subfigure}[b]{0.45\textwidth}
            \centering
            \includegraphics[width=\textwidth]{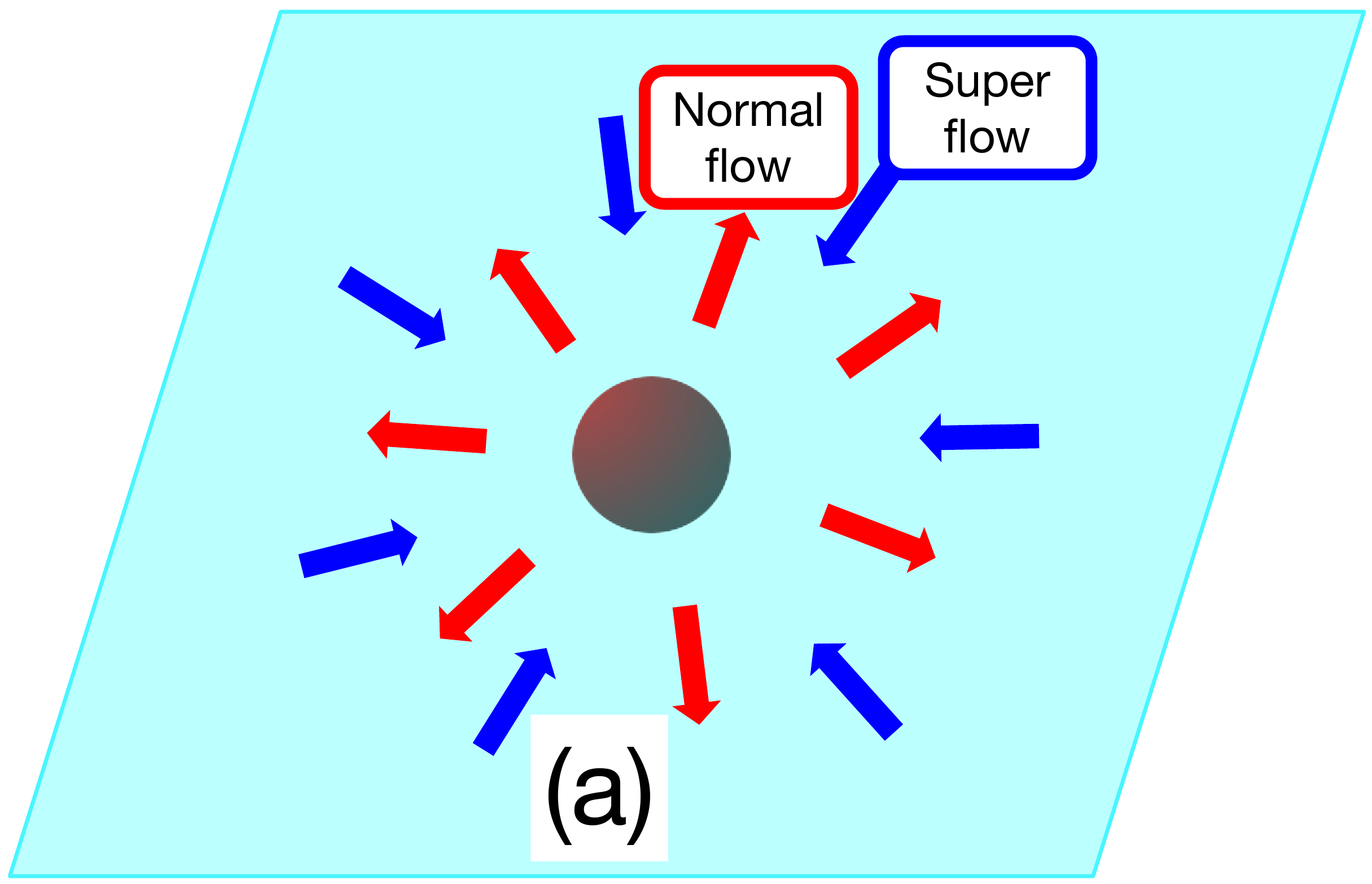}
    			\phantomcaption\label{fig:Radial_CF}
    \end{subfigure}
    \begin{subfigure}[b]{0.3\textwidth}
            \centering
            \includegraphics[width=\textwidth]{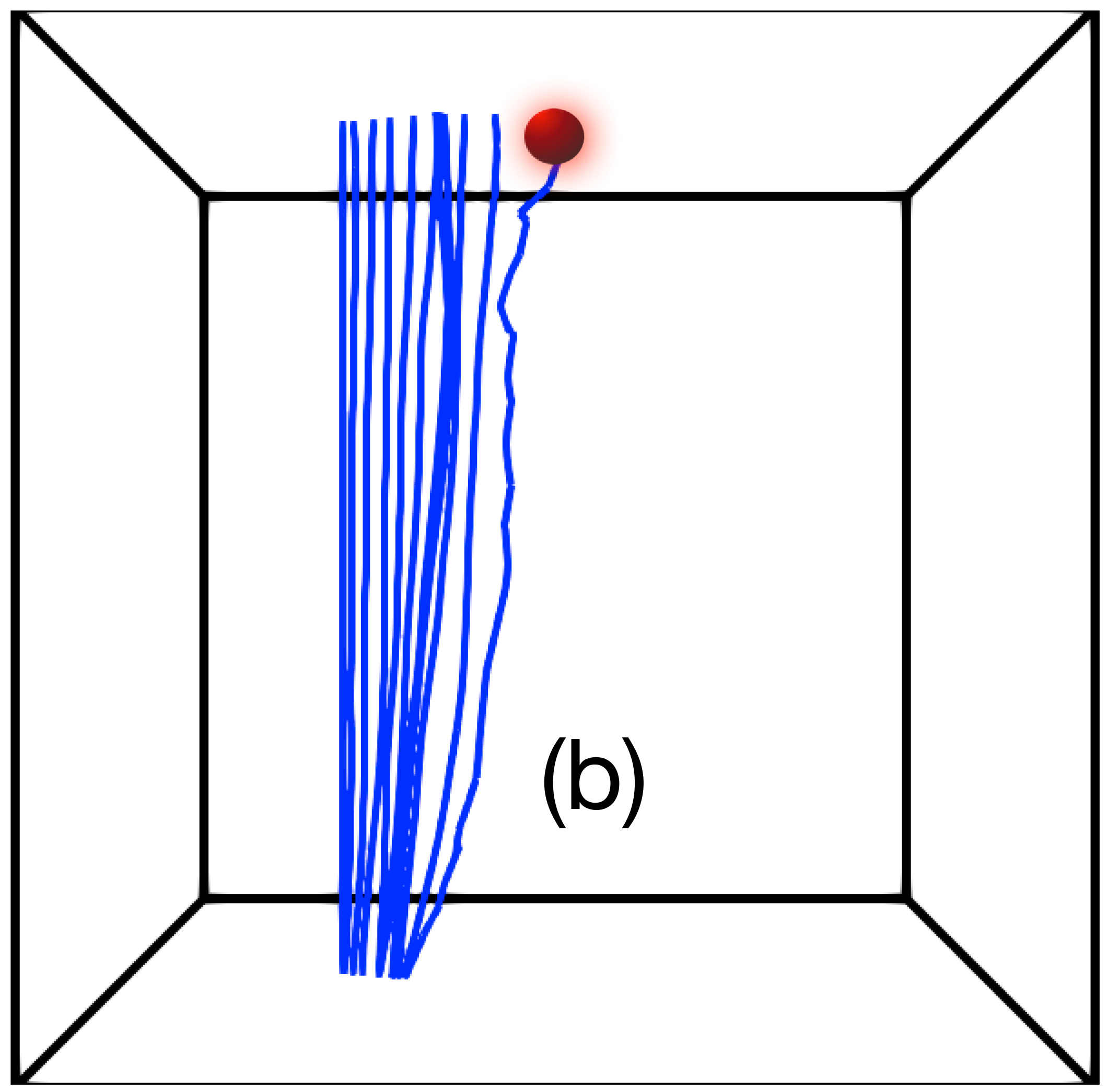}
    		   \phantomcaption\label{fig:Vortex_attraction}
    \end{subfigure}
    \caption{(a) Illustration of the local radial counter flow around a heated particle.  Normal flow is radiated away from the particle, while \red{superflow} is toward the particle.  The figure (b) shows how a particle of radius $5$ $\mu$m attracts  a vortex that is initially separated by a distance $50$ $\mu$m and reconnects with it in a cubic system of length $200$ $\mu$m.  \blue{Filament positions are shown with a time step of} $4\times10^{-3}$ s, and the whole process occurs within $0.04$ s.}
\end{figure}

If a particle is heated, it should establish a local radial counter flow as shown in Fig.\ref{fig:Radial_CF}, and the relative velocity \red{of normal and superfluid components} should be given by
\begin{equation}
v = \frac{Q}{\rho ST},\label{eq:CF}
\end{equation}
where $Q$ is the heat flux, $\rho$ is the density, $S$ is the entropy of the normal component, and $T$ is the temperature of the system.  In \blue{the} experiment \red{by Moroshkin \et.}, \blue{the largest possible heat flux at the particle surface illuminated by the laser is estimated as several W/cm$^2$. The counterflow velocity predicted by Eq. \eqref{eq:CF} for $Q=1$ W/cm$^2$ is $37$ cm/s}, which is significantly larger than the critical vortex tangle velocity ($\sim1$cm/s)\cite{Tough_1982}.  Even though the strength of the counter flow drops inversely proportional to the square of the distance from the center of the particle, the velocity is still greater than \blue{$1$ cm/s} within the spherical shell of radius $\sim 6.08 a_p$, where $a_p$ is the particle radius. 

 That could result in two effects; one is in the increase of effective particle radius by carrying both normal and superfluid with it, and the other is in attracting free quantized vortices toward it.  The latter effect is confirmed computationally by introducing a particle that sets up a steady local counter flow, neglecting the motion of the particle  for the simplicity.  

Figure $5(b)$ shows the case where the particle of radius \blue{$5$ $\mu$m emits a heat flow of $1$ W/cm$^2$}  and depicts how a vertical vortex filament initially placed $50$ $\mu$m away from the particle is moving toward \blue{it}.  This attraction seems to  enhance the possibility that a particle is trapped and exhibits the type (1) motion.  The origin of the type (2) motions, quasi-linear oscillations, is still under investigation, but the effects of particles creating local counter flows seem to be a clue to understand the phenomena.
\section{Conclusion and future work}
In summary, we have conducted calculations of the coupled dynamics of \red{a} quantized vortex and a fine particle to investigate the vortex-particle interactions at the surface of superfluid $^4$He based on the vortex filament model.  By considering a simple initial configuration, we can reproduce a motion that qualitatively explains the origin of the type (1) motion.  This scenario also suggests that the type (1) motion due to a particle whose mass density is less than that of solid \blue{helium} would be difficult to observe.  It is also confirmed that a heated particle is likely to attract free vortices through the local counter flow, which enhances the chance for a particle to be trapped by vortices and to exhibit the anomalous motions.

The analysis so far has been made mainly without treating the particle as local heaters.  In the future work we would like to investigate more closely the effects of the local counter flows and discuss how the motion of type (2) comes about.    
\label{sec:D}

%
%



\begin{thebibliography}{}
%
%
\bibitem{Moroshkin_2010}
P. Moroshkin, V. Lebedev, B. Grobety, C. Neururer, E. B. Gordon, and A. Weis, EPL. \B{90} 34002 (2010)

\bibitem{Moroshkin_2016}
P. Moroshkin, R.  Batulin, P. Leiderer, and K. Kono, Phys. Chem. Chem. Phys. \B{18}, 26444-26455 (2016)

\bibitem{Moroshkin_LT}
P. Moroshkin, P. Leiderer, K. Kono ``Motion of electrically charged metallic microparticles in superfluid helium", 28th International Conference on Low Temperature Physics, 9-16 August 2017, Goteborg, Sweden

\bibitem{Bewley_2006}
G.P. Bewley, D.P. Lathrop and K.R. Sreenivasan, Nature (London). \B{441}, 588 (2006)

\bibitem{Sergeev_2009}
\blue{Y. A. Sergeev, and C. F. Barenghi, J. Low Temp. Phys. \B{157}, 429-475 (2009)}

\bibitem{Guo_2014}
W. Gao, M. La Mantia, D. P. Lathrop, and S. W. Van Sciver, Proc. Natl. Acad. Sci. USA. \B{111}, 4653 (2014)

\bibitem{Moroshkin_2017}
P. Moroshkin, P. Leiderer, Th. B. M\"{o}ller, and K. Kono, Phys. Rev. E \B{95}, 053110 (2017)

\bibitem{Vinen_1995}
W. F. Vinen, Z. Phys. B \B{98}, 299-301 (1995)

\bibitem{Tabbert_1997}
\blue{B. Tabbert, H. G\"{u}nther, and G. zu Putlitz, J. Low temp. Phys. \B{109}, 653-707 (1997)}

\bibitem{Schwarz_1985}
K. W. Schwarz, Phys. Rev. \blue{B} \B{31}, 5782 (1985)

\bibitem{Schwarz_1988}
K. W. Schwarz, Phys. Rev. \blue{B} \B{38}, 2398 (1988)

\bibitem{Adachi_2010}
H. Adachi, S. Fujiyama, and M. Tsubota, Phys. Rev. B \B{81}, 104511 (2010)

\bibitem{Mineda_2013}
Y. Mineda, M. Tsubota, Y. A. Sergeev, C. F. Barenghi, and W. F. Vinen, Phys. Rev. B \B{87}, 174508 (2013)


\bibitem{Tough_1982}
J. T. Tough, Progress in Low Temperature Physics, vol. 8, 133-219. North-Holland, Amsterdam(1982)

\end{thebibliography}


\end{document}